\documentclass[times, twoside, watermark]{StyleBioRxiv}
\usepackage[utf8]{inputenc}
\usepackage{amsfonts,amsmath,amssymb}
\usepackage{array,graphicx}
\usepackage{tabularx,multirow}

\DeclareGraphicsExtensions{.pdf, .png}
\leadauthor{Irving-Pease} 
\begin{document}

\title{Quantitative Human Paleogenetics: what can ancient DNA tell us about complex trait evolution?}
\shorttitle{Quantitative Human Paleogenetics}

\author[1,\Letter]{Evan K. Irving-Pease}
\author[1]{Rasa Muktupavela}
\author[2]{Michael Dannemann}
\author[1,\Letter]{Fernando Racimo}

\affil[1]{Lundbeck Foundation GeoGenetics Centre, GLOBE Institute, University of Copenhagen, Copenhagen, Denmark}
\affil[2]{Center for Genomics, Evolution and Medicine, Institute of Genomics, University of Tartu, Tartu, Estonia}

\maketitle

\begin{abstract}
Genetic association data from national biobanks and large-scale association studies have provided new prospects for understanding the genetic evolution of complex traits and diseases in humans. In turn, genomes from ancient human archaeological remains are now easier than ever to obtain, and provide a direct window into changes in frequencies of trait-associated alleles in the past. This has generated a new wave of studies aiming to analyse the genetic component of traits in historic and prehistoric times using ancient DNA, and to determine whether any such traits were subject to natural selection. In humans, however, issues about the portability and robustness of complex trait inference across different populations are particularly concerning when predictions are extended to individuals that died thousands of years ago, and for which little, if any, phenotypic validation is possible. In this review, we discuss the advantages of incorporating ancient genomes into studies of trait-associated variants, the need for models that can better accommodate ancient genomes into quantitative genetic frameworks, and the existing limits to inferences about complex trait evolution, particularly with respect to past populations.
\end {abstract}

\begin{corrauthor}
evan.irvingpease\at gmail.com, \\
fernandoracimo\at gmail.com
\end{corrauthor}

\section*{1 - Complex trait genomics and ancient DNA}

The last decade has seen dramatic advances in our understanding of the genetic architecture of polygenic traits \citep{Visscher2017-nc}. The advent of genome-wide association studies (GWAS), with large sample sizes and deep phenotyping of individuals, has led to the identification of thousands of loci associated with complex traits and diseases \citep{Bycroft2018-ik,Buniello2019-ac,MacArthur2017-lz}. The resulting associations, and their inferred effect sizes, have enabled the development of so-called polygenic risk scores (PRS), which summarise either the additive genetic contribution of single nucleotide polymorphisms (SNPs) to a quantitative trait (e.g., height), or the increase in probability of a binary trait (e.g., major coronary heart disease) \citep{Dudbridge2013-vx}. For some well-characterised medical traits, like cardiovascular disease, the predictive value of PRS has led to their adoption in clinical settings \citep{Knowles2018-rr}; however, the accuracy of PRS remains limited to populations closely related to the original GWAS cohort \citep{Martin2019-qd} and can vary within populations due to age, sex and socioeconomic status \citep{Mostafavi2020-ly}. Ancient genomics has yielded considerable insights into natural selection on large-effect variants \citep{Malaspinas2016-ap,Dehasque2020-dl}, and an increasing number of studies are also now utilizing ancient genomes to learn about polygenic adaptation; the process by which natural selection acts on a trait with a large number of genetic loci, leading to changes in allele frequencies at many sites across the genome. Among these studies, the most commonly inferred complex traits are pigmentation and standing height.

Skin, hair and eye pigmentation are among the least polygenic complex traits; though more than a hundred pigmentation-associated loci have been found, their heritability is largely dominated by large-effect common SNPs \citep{Sturm2008-zc,Eiberg2008-xd,Sulem2007-fo,Han2008-ys,Liu2015-ws,OConnor2019-xp,Hider2013-jr}. Additionally, several of these variants have signatures of past selective sweeps detectable in present-day genomes \citep{Sabeti2007-tx,Lao2007-ea,Pickrell2009-ve,Rocha2020-he}. Nevertheless, genomic analyses in previously understudied populations—like sub-Saharan African groups—suggest that perhaps hundreds of skin pigmentation alleles of small effect remain to be found \citep{Martin2017-zm}. Similarly, recent studies have shown that eye pigmentation is far more polygenic than previous thought \citep{Simcoe2021-qf}. Recent quantitative and molecular genomic studies are painting an increasingly complex picture of the architecture of these traits, featuring more considerable roles for epistasis, pleiotropy and small-effect variants than were previously assumed (for an extensive review of skin pigmentation, see \cite{Quillen2019-fu}).

Recently, ancient DNA (aDNA) studies have attempted to reconstruct pigmentation phenotypes in ancient human populations, although the extent to which these predictions are accurate remains uncertain. These reconstructions have been mostly focused on ancient individuals from Western Eurasia, due to the relatively higher abundance of SNP-phenotype associations from European-centric studies, and the poor portability of gene-trait associations to more distantly related populations \citep{Martin2017-kp,Martin2019-qd}. For example,  \cite{Olalde2014-wy} queried pigmentation-associated SNPs in genomes of Mesolithic hunter-gatherer remains from western and central Eurasia, and suggested that the lighter skin colour characteristic of Europeans today was not widely present in the continent before the Neolithic. \cite{Gonzalez-Fortes2017-di} analysed Mesolithic and Eneolithic genomes from central Europe, and inferred dark hair, brown eyes and dark skin pigmentation for the Mesolithic individuals and dark hair, light eyes, and lighter skin pigmentation for an Eneolithic individual. Similarly, \cite{Brace2019-yh} inferred pigmentation phenotypes for Mesolithic and Neolithic genomes from western Europe, and reported that the so-called ‘Cheddar Man’, a Mesolithic individual from present-day England, had blue/green eyes and dark to black skin, in contrast to later Neolithic individuals with dark to intermediate skin pigmentation. Contrastingly, \cite{Gunther2018-ic} found elevated frequencies of light skin pigmentation alleles in individuals from the Scandinavian Mesolithic, suggestive of early environmental adaptation to life at higher latitudes. These reconstructions have also been carried out in individuals with no skeletal remains; for example, \cite{Jensen2019-zi} used pigmentation-associated SNPs to infer the skin, hair and eye colour of a female individual whose DNA was preserved in a piece of birch tar ‘chewing gum’.

Some aDNA studies have sought to systematically investigate how pigmentation-associated variants were introduced and evolved in the European continent. \cite{Wilde2014-sm} was one of the first studies to provide aDNA-based evidence that skin, hair, and eye pigmentation-associated alleles have been under strong positive selection in Europe over the past 5,000 years. The first large-scale population genomic studies \citep{Haak2015-zl,Mathieson2015-xg,Allentoft2015-pa} showed that major effect alleles associated with light eye colour likely rose in frequency in Europe before alleles associated with light skin pigmentation. More recently, \cite{Ju2021-dj} argued that the increase in light skin pigmentation in Europeans was primarily driven by strong selection at a small proportion of pigmentation-associated loci with large effect sizes. When testing for polygenic adaptation using an aggregation of all known pigmentation-associated variants, they did not detect a statistically significant signature of selection.

The other trait that has shared comparable prominence with pigmentation in the aDNA literature is standing height. In contrast to pigmentation, the genetic architecture of height is highly polygenic \citep{Yang2015-cs,Yengo2018-zd,Bycroft2018-ik}. The heritability of this trait is dominated by a large number of alleles with small effect sizes, and shows strong evidence for negative selection in present-day populations \citep{OConnor2019-xp}. Studies of the genetic component of height in ancient populations have shown that ancient West Eurasian populations were, on average, more highly differentiated for this trait than present-day West Eurasian populations, and more so than one would predict from genetic drift alone \citep{Mathieson2015-xg,Cox2019-ae,Martiniano2017-ke}. \cite{Cox2019-ae} compared predicted genetic changes in height in ancient populations to inferred height changes estimated via skeletal remains. They concluded that the changes in inferred standing height were partially predicted by genetics; with both measures remaining relatively constant between the Mesolithic and Neolithic, and increasing between the Neolithic and Bronze Age. A follow-up study by \cite{Cox2021-yw} used polygenic scores for height to show that PRS predicts 6.8\% of the observed variance in femur length in ancient skeletons, after controlling for other variables. This is approximately one quarter of the predictive accuracy of PRS in present-day populations; which the authors attribute to the low-coverage aDNA data used in their study. Contrastingly, \cite{Marciniak2021-nt} used the discordance between PRS for height, calculated from aDNA, and height inferred from the corresponding skeletal remains, to argue that Neolithic individuals were shorter than expected due to either poorer nutrition or increased disease burden, relative to hunter-gatherer populations.

However, the inference of standing height from skeletal remains is not without its own problems. Both \cite{Cox2021-yw} and \cite{Marciniak2021-nt} used the method developed by \cite{Ruff2012-ar} to estimate stature from skeletal remains. Nevertheless, their respective estimates of stature—based on femur length—varied between some of the individuals included in both studies. Where multiple skeletal elements were available for ancient individuals, \cite{Marciniak2021-nt} also produced separate stature estimates from femur, tibia, humerus and radius length, which varied substantially within some individuals; highlighting the uncertainty in estimates of stature from skeletal remains.

\begin{figure*}
\centering
\includegraphics[width=1\linewidth]{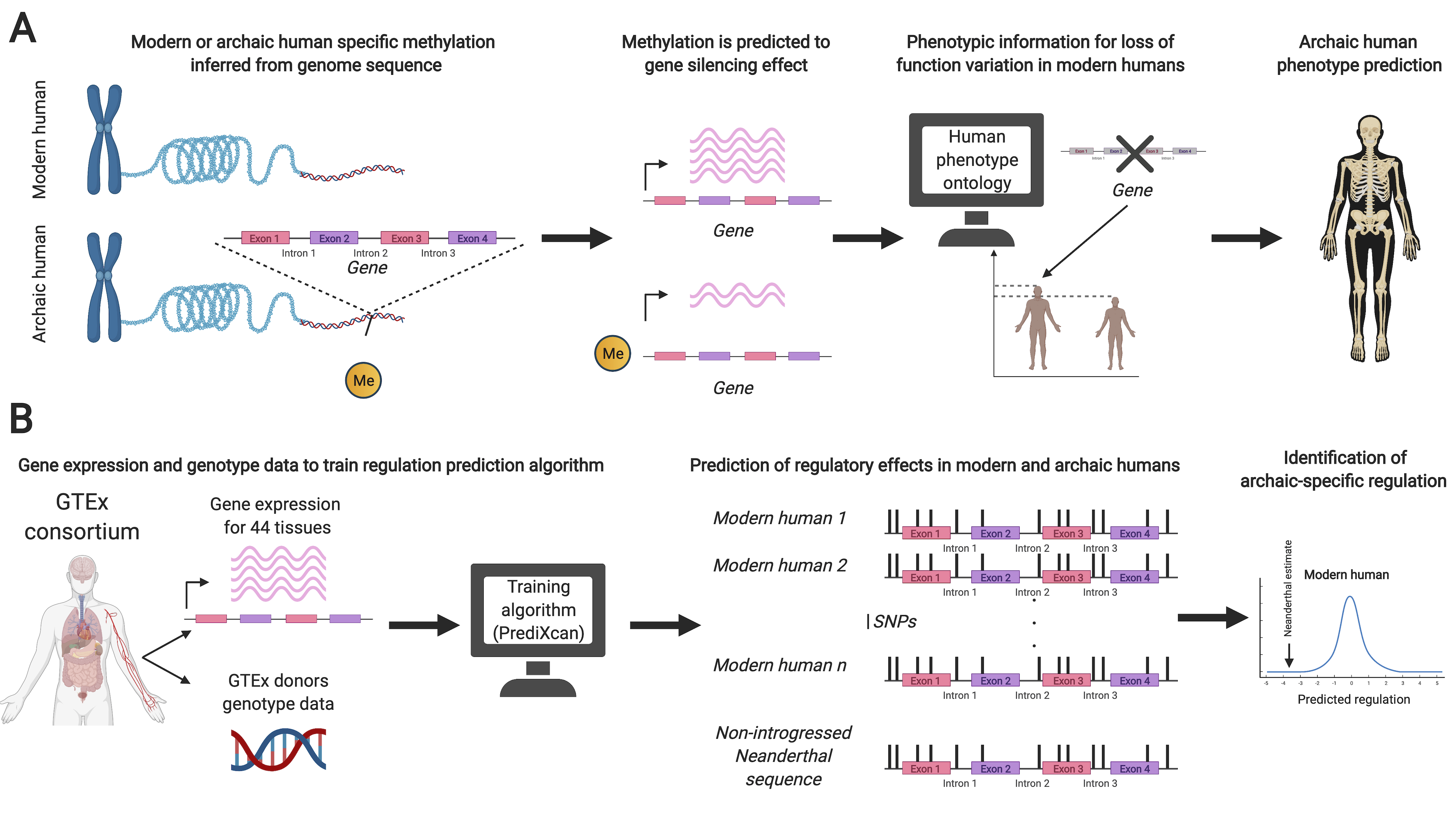}
\caption{A) Schematic illustration of the prediction method used by \cite{Gokhman2020-am} to infer archaic human phenotypes based on methylation maps. B) Schematic illustration of the method by \cite{Colbran2019-fu} to predict regulatory effects of non-introgressed archaic human DNA.}
\label{fig:1}
\end{figure*}

\section*{2 - Inferring complex traits in archaic hominids}

The availability of genome sequences from archaic humans, like Neanderthals and Denisovans, has greatly expanded our understanding of their demographic history and interactions with modern humans \citep{Prufer2017-eo,Prufer2014-oq,Meyer2012-xm}. However, little is known about complex traits in archaic humans, besides what can be inferred directly from their skeletal remains. In the case of Denisovans, such remains are presently limited to a few teeth, a mandible and other small bone fragments, making it difficult to make confident inferences of their biology \citep{Chen2019-ig,Meyer2012-xm,Sawyer2015-tf,Slon2017-ii}. However, past admixture events with archaic human groups have left a genetic legacy in present-day people, providing a possible inroad to study archaic human biology \citep{Sankararaman2012-mv}. Today, around 2\% of the genomes of non-African humans are known to be descended from Neandertals, and an additional ~5\% of the genomes of people in Oceania can be traced back to Denisovans \citep{Sankararaman2016-ac,Sankararaman2014-bx,Vernot2016-gl,Vernot2014-yh}. 

Knowledge about admixture between archaic and modern humans has led to a recent flurry of exploratory studies concerning the potential impact of archaic variants on complex traits in present-day populations. Various approaches have been used to identify introgressed archaic DNA putatively under positive selection in modern humans \citep{Gittelman2016-sc,Sankararaman2016-ac,Sankararaman2014-bx,Vernot2014-yh,Vernot2016-gl,Racimo2017-eu,Khrameeva2014-ec,Perry2015-cn}. Overall, these studies have shown that archaic DNA is linked to pathways related to metabolism, as well as skin and hair morphology. Via association studies, Neanderthal variants in specific loci have been shown to influence several disease and immune traits, as well as skin and hair color, behavioural traits, skull shape, pain perception and reproduction \citep{Skov2020-eu,Dannemann2016-sc,Sams2016-jz,Zeberg2020-me,Zeberg2020-xe,Zeberg2020-lz,Gunz2019-va,Sankararaman2014-bx,Zeberg2021-kq}. 

Additionally, comparisons between the combined phenotypic effects of Neandertal variants and frequency-matched non-archaic variants have revealed that Neanderthal DNA is over-proportionally associated with neurological and behavioural phenotypes, as well as viral immune responses and type 2 diabetes \citep{Simonti2016-md,Dannemann2017-gq,Dannemann2021-pd,Quach2016-tj}. These groups of phenotypes may be linked to environmental factors, such as ultraviolet light exposure, pathogen prevalence and climate, that substantially differed between Africa and Eurasia. It has been suggested that the over-proportional contribution of Neandertal DNA to immunity and behavioural traits in present-day humans might be a reflection of adaptive processes in Neandertals to these environmental differences. In comparison, much less is known about the impact of Denisovan DNA on complex traits, because limited phenotypic data are presently available from present-day populations. However, individual Denisovan-like haplotypes found in high frequencies in some human populations have been associated with high altitude adaptation and fat metabolism \citep{Racimo2017-kw,Huerta-Sanchez2014-or}. 

One key limitation to these approaches is that only about 40–50\% of the Neandertal genome can be recovered in present-day humans, and therefore discoverable in such analyses \citep{Vernot2014-yh,Sankararaman2014-bx,Skov2020-eu}. Furthermore, the majority of tested cohorts used for such studies are of European ancestry, which limits analyses to archaic variants present in these populations. This is particularly notable since Neandertal phenotype associations in European and Asian populations have been shown to contain population-specific archaic variants \citep{Dannemann2021-pd}. It has also been shown that negative selection, soon after admixture, has played an important role in removing some of the missing segments of archaic DNA \citep{Harris2016-iu,Juric2016-kw,Petr2019-tb}. It is therefore possible that missing segments of archaic DNA had strong phenotypic effects. For archaic DNA that does persist in present-day populations, much of it is segregating at low allele frequencies, making it difficult to confidently link it to phenotypic effects. 

Furthermore, it remains questionable how transferable any phenotypic associations are between modern and archaic humans, given the difficulties of transferring associations between present-day populations \citep{Duncan2019-sj,Martin2017-kp}. All of the above studies have used gene-trait association information from analyses carried out in modern humans. It remains undetermined if the phenotypic effects of archaic DNA in present-day populations are a reliable proxy for phenotypic effects in archaic humans themselves.

Recent studies have also aimed to predict the phenotypic effects of archaic DNA without relying on introgression in present-day populations (see Figure \ref{fig:1}). \cite{Colbran2019-fu} used a machine learning algorithm, trained on genetic variation in present-day humans, to infer putative regulatory effects on variation present only in Neandertal genomes. \cite{Gokhman2020-am,Gokhman2020-ej} used aDNA damage patterns to infer a DNA methylation map of the Denisovan genome, and linked the inferred regulatory patterns to loss-of-function phenotypes, in order to predict their skeletal morphology and vocal and facial anatomy. It remains to be seen how successful these approaches are at predicting archaic human phenotypes. A possible inroad into validation could rest on functional assays for testing and evaluating the phenotypic impact of archaic DNA \citep{Dannemann2020-yc,Trujillo2021-pg,Dannemann2021-qi}.

\section*{3 - The challenge of detecting polygenic adaptation in ancient populations}

Perhaps the most fascinating question about the evolution of complex traits in humans is whether they were subject to natural selection. Current methods to detect polygenic adaptation have mainly focused on present-day populations; using either differences between populations, or variation within them, to identify polygenic adaptation. For example, \cite{Berg2014-mw} developed a method that identifies over-dispersion of genetic values among populations, compared to a null distribution expected under a model of drift; which \cite{Racimo2018-ww} extended to work with admixture graphs. \cite{Field2016-qm} used the distribution of singletons around trait-associated SNPs, and \cite{Uricchio2019-hy} used the joint distribution of variant effect sizes and derived allele frequencies (DAF). Whichever method is used, significant caveats must be addressed before attributing differences in such scores to polygenic adaptation \citep{Novembre2018-vo,Rosenberg2019-xz,Coop2019-vo}. Most of these issues affect both present-day and ancient populations, but many are especially problematic when working with ancient genomes.

\begin{figure}[t]
\begin{center}
\includegraphics[height=0.5\textheight]{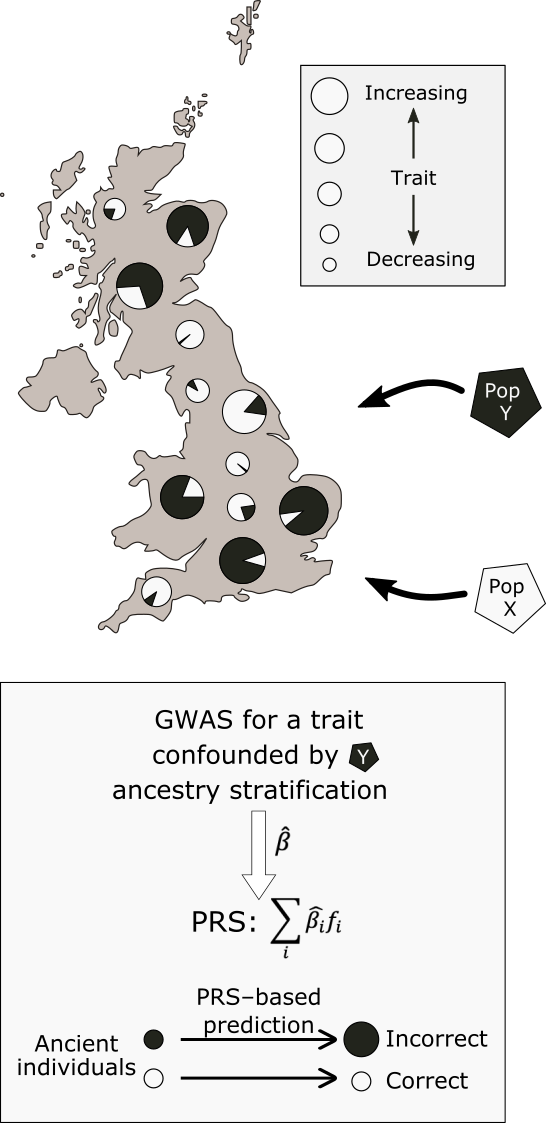}
\end{center}
\caption{Potential effect of population stratification in GWAS. Two ancestral populations, X and Y, have contributed differing ancestry proportions to present-day individuals. Due to non-genetic environmental effects, individuals with a larger proportion of population Y ancestry have higher values for a measured trait. This may lead to biased GWAS effect size estimates, which associate population Y ancestry with increasing values of the trait. When used to make inferences about the past, this would lead to systematically inflated polygenic scores for this trait in samples from population Y.}%
\label{fig:2}
\end{figure}

A prominently reported example of polygenic adaptation is that of selection for increasing height across a north-south gradient in Europe \citep{Turchin2012-rs,Berg2014-mw,Robinson2015-xv,Zoledziewska2015-qj,Berg2019-kp,Racimo2018-ww,Guo2018-or,Chen2020-bk}. Most studies which described this signal based their analyses on effect size estimates from the GIANT consortium, a GWAS meta-analysis encompassing 79 separate studies \citep{Wood2014-tj}. Concerningly, follow-up work using the larger and more homogeneous UK Biobank cohort failed to replicate the signal of polygenic adaptation for height \citep{Berg2019-oc,Sohail2019-zd}. A recent systematic comparison across a range of GWAS cohorts has further shown that the results of these tests are highly dependent on the ancestry composition of the cohort used to obtain the effect size estimates \citep{Refoyo-Martinez2021-kb}. These analyses showed that residual stratification in GWAS meta- and mega-analyses can result in inflated effect size estimates that, in turn, can lead to spurious signals of selection. The effects of this residual stratification may be exacerbated for ancient populations with non-uniform relatedness to present-day GWAS cohorts (see Figure \ref{fig:2}).

Residual stratification is a major concern for GWAS, even among a relatively homogeneous cohort like the UK Biobank. \cite{Zaidi2020-xb} used simulations to show that fine-scale recent demography can confound GWAS which has been corrected for stratification using common variants only. Failure to adequately correct for localised population structure can lead to spurious associations between a trait and low-frequency variants that happen to be common in areas of atypical environmental effect. This finding is problematic as most GWAS have been conducted on either SNP array data, or on genomes imputed from SNP array data \citep{Visscher2017-nc}. For example, GWAS summary statistics from the UK Biobank are based on imputed genomes \citep{Bycroft2018-ik}. A limitation of this approach is that the accuracy of imputed genotypes are inversely correlated with the minor allele frequencies (MAF) of variants in the reference panel. Additionally, rare variants that are not segregating in the reference panel cannot be imputed at all. As a result, imputed genomes are specifically depleted in the rare variants needed to adjust for stratification from recent demography. 

For large sample sizes, low-frequency variants (MAF $\leq$ 0.05) make a significant contribution to the heritability of many complex traits \citep{Hartman2019-dr,Mancuso2016-zc}, but the role of rare variants is less well established. Both empirical and simulation studies have shown that for traits under either negative or stabilising selection, there is an inverse correlation between effect size and MAF \citep{Durvasula2021-qd,Schoech2019-ab,Simons2018-eg}. For the many traits thought to be under negative selection \citep{OConnor2019-xp}, large effect variants that are rare in present-day populations may have had higher allele frequencies in ancient populations due to selection. This makes polygenic scores for ancient individuals especially sensitive to bias from GWAS effect size estimates ascertained from common variants only. Conversely, where present-day rare variants with large-effect sizes are known, higher frequencies in ancient populations would result in more accurate PRS predictions, due to their larger contribution to the overall genetic variance.

A recent analysis indicated that a substantial component of the unidentified heritability for anthropometric traits like height and BMI lies within large effect rare variants, some with MAF as low as 0.01\% \citep{Wainschtein2019-ir}. However, using GWAS to recover variant associations for SNPs as rare as this would require hundreds of thousands of whole-genomes, substantially exceeding the largest whole-genome GWAS published to date (e.g., \cite{Taliun2021-ls}). The consequence of this missing heritability may be particularly acute for trait prediction in ancient samples, as large-effect rare variants which contributed to variability in the past may no longer be segregating in present-day populations. Indeed, simulations suggest that the genetic architecture of complex traits is highly specific to each population, and that negative selection enriches for private variants, which contribute to a substantial component of the heritability of each trait \citep{Durvasula2021-qd}. Empirical studies have also identified that functionally important regions, including conserved and regulatory regions, are enriched for population-specific effect sizes, and that this pattern may have been driven by directional selection \citep{Shi2021-au}.

In addition to these issues, the majority of SNP associations inferred from GWAS are not the causal alleles. Instead, GWAS predominantly identifies SNPs which are in high linkage disequilibrium (LD) with causal alleles. Most GWAS also assume a model in which all complex trait heritability is additive and well tagged by SNPs segregating in the cohort; although some GWAS do include non-additive models (e.g., \cite{Guindo-Martinez2021-ei}). Consequently, effect size estimates are contingent on the LD structure of the cohort in which they were ascertained. Due to recombination, this LD structure decays through time, and is reshaped by the population history in which selection processes are embedded. 

Over the last decade, paleogenomic studies have repeatedly demonstrated that the evolutionary histories of human populations are characterized by recurrent episodes of divergence, expansion, migration and admixture (reviewed in \cite{Pickrell2014-pt,Skoglund2018-bc}). For example, in West Eurasia, four major ancestry groups have contributed to the majority of present-day genetic variability \citep{Jones2015-jr}. As such, the LD structure of present-day British individuals—which underpins effect size estimates from the UK Biobank—was substantially different prior to the Bronze Age, when the most recent of these major admixture episodes occurred \citep{Haak2015-zl,Allentoft2015-pa}. To improve ancestral trait prediction, new methods which explicitly model the haplotype structure of both ancient populations and present-day GWAS cohorts are needed.

In aggregate, these issues combine to substantially diminish the portability of polygenic scores between populations. Indeed, in present-day populations, the predictive accuracy of PRS degrades approximately linearly with increasing genetic distance from the cohort used to ascertain the GWAS \citep{Scutari2016-vb,Martin2017-kp,Kim2018-pi,Martin2019-qd,Mostafavi2020-ly,Bitarello2020-yf,Majara2021-nb}. Even within a single ancestry group, the correlation between PRS calculated from different discovery GWAS shows considerable variance \citep{Schultz2021-lo}. However, the extent to which the issue of PRS portability also affects ancient populations, which are either partially or directly ancestral to the GWAS cohort, are yet to be determined.

In cases where a robust signal of polygenic adaptation can be identified, care must still be taken when interpreting which trait was actually subject to directional selection. Due to the highly polygenic nature of most complex traits, there is a high rate of genetic correlation between phenotypes \citep{Shi2017-hu,Ning2020-vg}. This can occur when correlated traits share causal alleles (i.e., pleiotropy) or where casual alleles are in high LD with each other. Consequently, selection acting on one specific trait can generate a spurious signal of polygenic adaptation for multiple genetically correlated traits. Recently, \cite{Stern2021-ak} developed a method for conditional testing of polygenic adaptation to address this problem. When considered in a joint test, previously identified signals of selection for educational attainment and hair colour in British individuals were significantly attenuated by the signal of selection for skin pigmentation \citep{Stern2021-ak}. However, this approach can only untangle genetic correlations between traits which have been measured in GWAS cohorts, leaving open the possibility that selection is acting on an unobserved yet correlated trait. Indeed, many GWAS traits are either coarse proxy measures with substantial socio-economic confounding (e.g., educational attainment), or narrow physiological measurements (e.g., levels of potassium in urine); neither of which are likely to have been direct targets of polygenic adaptation. In practice, the truly adaptive phenotype is rarely directly observable, and all measured traits are genetically correlated proxies at various levels of abstraction.

\section*{4 - Limitations and caveats specific to ancient DNA}

In addition to all of the general issues and caveats discussed above, working with ancient DNA also involves a range of issues that are particular to the degraded nature of the data; such as post-mortem damage, generally low average sequence coverage, short fragment lengths, reference bias, and microbial and human contamination \citep{Dabney2013-ew,Peyregne2020-kg,Gilbert2005-lf,Renaud2019-vj}. All of these factors affect our ability to correctly infer ancient genotypes; and therefore, to construct accurate polygenic scores or infer polygenic adaptation. 

A common strategy for dealing with the low endogenous fraction of aDNA libraries is to use in-solution hybridisation capture to retrieve specific loci, or a set of predetermined SNPs \citep{Cruz-Davalos2017-cz,Avila-Arcos2011-zu}. This approach has substantial advantages in on-target efficiency, at the cost of ascertainment bias. For example, in the case of the popular ‘1240k’ capture array, targeted SNPs were predominantly ascertained in present-day individuals \citep{Haak2015-zl,Fu2015-lz}. Consequently, an unknown fraction of the true ancestral variability is lost during capture. This is further exacerbated by the generally low coverage of most aDNA libraries; for which a common practice is to draw a read at random along each position in the genome, to infer ‘pseudo-haploid’ genotypes. When used to compute polygenic scores for ancient populations, only a subset of GWAS variants can be used, which substantially reduces predictive accuracy. \cite{Cox2021-yw} estimate that the combined effect of low-coverage and pseudo-haploid genotypes reduced their predictive accuracy by approximately 75\%, when compared to present-day data.

An alternative approach is to perform low-coverage shotgun sequencing, followed by imputation, using a large reference panel \citep{Ausmees2019-gk,Hui2020-as}. This has the dual advantages of reducing ascertainment bias and increasing the number of GWAS variants available to calculate polygenic scores. However, imputation itself introduces a new source of bias, particularly if the reference panel is not representative of the ancestries found in the low-coverage samples. Nevertheless, the level of imputation bias can be empirically estimated by downsampling high-coverage aDNA libraries and testing imputed genotypes against direct observations (e.g., \cite{Margaryan2020-wy}). Where a suitable reference panel exists, recently developed methods for imputation from low-coverage sequencing data \citep{Rubinacci2021-ip,Davies2021-zc} show great promise for ancient DNA studies (e.g., \cite{Clemente2021-ey}).

Even under ideal conditions, in which exact polygenic scores for ancient populations are known \textit{a priori}, interpreting differences in mean PRS between groups still requires careful consideration. For many polygenic traits, the variance between population means is lower than the variance within populations. As a result, differences in population level polygenic scores have limited predictive value for inferring the physiology or behaviour of individual people in the past. Genetics plays only a partial role in shaping phenotypic diversity, and differences in polygenic scores between individuals, or populations, does not automatically translate into differences in the expressed phenotype. Indeed, for some complex traits, an inverse correlation has been observed; in which polygenic scores have been steadily decreasing over recent decades, whilst the measured phenotype has been increasing (e.g., educational attainment \citep{Kong2017-kf,Abdellaoui2019-oo}). This highlights the substantial role of environmental variation in shaping phenotypic diversity. For ancient populations, we must also consider the wide variation in culture, diet, health, social organisation and climate which will have mediated any potential differences in population level polygenic scores. Furthermore, ancient populations are likely to have experienced a heterogeneous range of selective pressures. What we observe in present-day populations is not the result of a single directional process, but instead represents a mosaic of haplotypes which were shaped by different fitness landscapes, at varying levels of temporal depth.

Lastly, in most cases, we cannot directly observe phenotypes in the ancient individuals whose genomes have been studied. This greatly limits our ability to compare the genetically predicted value of a trait to its expressed phenotype, raising the question: are predictions of most ancient phenotypes inherently unverifiable? For well-preserved traits, like standing height, there is considerable variability in estimates produced from different skeletal elements and between different studies \citep{Cox2021-yw,Marciniak2021-nt}. For traits that do not preserve well in the archaeological record, the prospects of validation are much poorer. These include not only soft tissue measurements (e.g., pigmentation or haemoglobin counts), but also personality and mental health traits that require an individual to be alive to be properly measured or diagnosed. Furthermore, some phenotypes are nonsensical outside of a modern context. Whilst it is possible to build a polygenic score for “time spent watching television” (UK Biobank code: 1070), it is not clear how to interpret any potential differences one might find between Mesolithic hunter-gatherers and Neolithic farmers. This problem extends more generally to all phenotypes which have strong gene–environment interactions, in which the expression of the trait may have been substantively different in the past due to diverse environmental conditions (e.g., the interplay between BMI and diet).

\section*{5 - Prospects for the future}

The growth in the number of ancient genomes currently shows little signs of slowing, nor does the increasing availability of gene-trait association data. Predictably, efforts to perform trait predictions in ancient individuals will also continue to grow. We believe that increased emphasis on limitations and caveats in the way we study and communicate these findings will enable a better understanding of what we can and cannot predict with existing models. 

As a working assumption, polygenic scores from any single GWAS should be considered unreliable in an ancient trait reconstruction analysis. Researchers should only trust observed signals of trait evolution if those patterns hold across multiple independent GWAS (e.g., \cite{Chen2020-bk}), and preferably where each of these GWAS has been performed on a large cohort with homogeneous ancestry \citep{Refoyo-Martinez2021-kb}.

We also need to better understand how well GWAS effect size estimates, ascertained in present-day populations, generalise to ancient populations that are only partially ancestral to the GWAS cohort. One approach to this would be to use simulations, under a plausible demographic scenario, to explore how the predictive accuracy of PRS degrades through time and across the boundaries of major ancestral migrations.

Traits that are preserved in the fossil record can provide a degree of partial benchmarking \citep{Cox2019-ae,Cox2021-yw}; however, the genetic components of variation are often only partially explained by polygenic scores, and environmental components almost always play large roles in expressed trait variation, often dwarfing the contribution of polygenic scores. Furthermore, only a few—largely osteological—traits are well preserved over time, so these comparisons will always be limited in scope.

That being said, there are several promising avenues of research that could serve to improve genetic trait prediction in ancient populations. An existing approach to improve the portability of PRS across ancestries is to prioritise variants with predicted functional roles \citep{Amariuta2020-gi,Weissbrod2020-jw}. This approach aims to improve PRS portability in present-day populations by reducing the fraction of spurious associations due to the cohort specific LD structure of the GWAS reference panel. Another promising approach is to jointly model PRS using GWAS summary statistics from multiple populations \citep{Marquez-Luna2017-fa,Ruan2021-jj,Turley2021-pw}. By including information from genetically distant groups, these methods can account for the variance in effect sizes inferred between GWAS cohorts. This multi-ancestry approach holds particular promise for ancient populations, as it may help to identify variant associations which are segregating in only a subset of present-day populations, but which were more widespread in the past.

These studies also underscore the importance of studying the ancestral haplotype backgrounds on which beneficial, deleterious or neutral alleles spread. Recent studies have shown that tests of selection on individual loci can gain power by explicitly modelling patterns of ancestry across the genome \citep{Pierron2018-xk,Hamid2021-if}. Strong selective signals might be masked by post-selection admixture processes, but might become evident once the ancestry of the selected haplotypes is explicitly modelled \citep{Souilmi2020-dv}. This phenomenon is also likely to affect polygenic adaptation studies, particularly when the degree of correlation between genetic score differences and differences in ancestral haplotype backgrounds is expected to be high, for example, after admixture between populations that have been evolving in isolation for long periods of time.

A promising avenue of research is developing around new methods for approximately inferring ancestral recombination graphs (ARG) \citep{Kelleher2019-yu,Speidel2019-zs}, which have recently been extended to incorporate non-contemporaneous sampling \citep{Wohns2021-iq,Speidel2021-sh}. An ARG is a data structure which contains a detailed description of the genealogical relationships in a set of samples, including the full history of gene trees, ancestral haplotypes and recombination events which relate the samples to each other at every site in the genome \citep{Griffiths1997-ev}. One potential advantage of such a data structure is that it may be used to help mitigate issues with the portability of polygenic scores. By building an ARG composed of both ancient samples and the present-day cohorts used to ascertain the GWAS associations, one could potentially determine which haplotypes are shared between the GWAS cohort and the ancient populations; thereby reducing effect size bias in populations that are only partially ancestral to the GWAS cohort.

Another area in which ancient genomes offer unique potential is in detecting polygenic adaptation in response to environmental change. The time-series nature of ancient genomes provides the potential for the incorporation of paleoclimate reconstructions (e.g., \cite{Brown2018-dh}) into tests of polygenic adaptation, in a manner that is not possible with present-day data alone.

Ultimately, the ancient genomics community must come to terms with the limitations of genetic hindcasting. Ancient genomes provide an unprecedented window into our past, but this window is often blurry and distorted. There is still a lot of information waiting to be obtained from ancient DNA, and some of the blurriness might ultimately come into focus as computational methods continue to improve. But we must also accept the fact that many aspects of past human biology—including physical characteristics and disease susceptibility—might be irrevocably lost to the tides of history.  Ancient genome sequences are, after all, molecular fossils: imperfect and degraded records of lives that ceased to exist long ago.

\begin{acknowledgements}
We thank the members of the Racimo group for helpful advice and discussions. EIP was supported by the Lundbeck Foundation (grant R302-2018-2155) and the Novo Nordisk Foundation (grant NNF18SA0035006). FR and RM were supported by a Villum Fonden Young Investigator award to FR (project no. 00025300). Additionally, FR was supported by the COREX ERC Synergy grant (ID 951385). MD was supported by the European Union through the Horizon 2020 Research and Innovation Programme under Grant No 810645 and the European Regional Development Fund Project No. MOBEC008.  Figure 1 was created with Biorender.com.
\end{acknowledgements}

\bibliography{references}


\newpage


\end{document}